\documentclass[aps,prl,groupedaddress,showpacs,twocolumn]{revtex4}
\usepackage{graphicx}
\usepackage{amssymb}
\def\be{\begin{equation}}
\def\ee{\end{equation}}

\def\vn{{\bf n}}
\def\vv{{\bf v}}
\def\vp{{\bf p}}
\def\vz{{\bf z}}
\newcommand{\vsigma}{\mbox{\boldmath$\sigma$}}
\def\Tr#1{\mbox{Tr}\left[#1\right]}

\newcommand{\g}{\mathfrak{g}}

\newcommand{\Area}{{\cal A}}

\begin{document}
\title{Dynamics of spin transport in voltage-biased Josephson junctions}
\author{Erhai Zhao}
\author{J. A. Sauls}
\affiliation{Department of Physics \& Astronomy, Northwestern
             University, Evanston, IL 60208}
\date{\today}
\pacs{72.25.-b, 72.25.Mk, 74.50.+r}
\begin{abstract}
We investigate spin transport in voltage-biased spin-active
Josephson junctions. The interplay of spin filtering, spin mixing,
and multiple Andreev reflection leads to nonlinear voltage
dependence of the \emph{dc} and \emph{ac} spin current. We compute
the voltage characteristics of the spin current ($I_{\textsf{\tiny
S}}$) for superconductor-ferromagnet-superconductor (\textsf{SFS})
Josephson junctions. The sub-harmonic gap structure of
$I_{\textsf{\tiny S}}(V)$ is shown to be sensitive to the degree of
spin mixing generated by the ferromagnetic interface, and exhibits a
pronounced even-odd effect associated with spin transport during
multiple Andreev reflection processes. For strong spin mixing both
the magnitude and the direction of the \emph{dc} spin current can be
sensitively controlled by the bias voltage.
\end{abstract}
\maketitle

By controlling the quasiparticle spin degree of freedom in
solid-state heterostructures, spintronics provides a new paradigm
and possibilities for making devices and circuits with improved functionality over
conventional electronics \cite{zut04}. Although spintronics research
has been mostly focused on hybrid structures of normal metal (N),
semiconductors, and ferromagnetic materials, superconductors has
increasingly proved to be a useful component in constructing
spintronic devices. This is because the proximity coupling of
superconductor (S) to ferromagnet (F), on the scale ranging from
nanometers to microns, offers a unique opportunity to probe the spin
degree of freedom and coherently control the spin polarized
current \cite{buz05,ber05}. For example, two leading methods of
measuring the spin polarization of ferromagnetic materials,
spin-dependent tunnelling \cite{ted94} and point contact Andreev
reflection spectroscopy \cite{sou98}, both rely on measuring the
conductance of superconductor-ferromagnet junctions \cite{valls,zut04}.
Recently spin-triplet supercurrent was created to flow through the
superconductor-ferromagnet-superconductor (SFS) Josephson junctions
\cite{kei06}. The spin current transported through in SFNFS structures was also
predicted to have properties with the potential of making
superconducting spin-transfer devices \cite{wai02,bel04}.

The equilibrium properties of SF hybrid systems have been
extensively studied in recent years \cite{buz05,ber05}. However, nonequilibrium 
spin-transport properties, which are essential for the purpose of
manipulating spin current flow, are far from being well understood. As an
example, the SFS $\pi$ junction has been proposed for constructing an
environmentally decoupled quantum bit \cite{iof99}. Optimizing its
performance requires knowledge of its \textit{dissipative} dynamics
under nonequilibrium conditions, such as the current-voltage
characteristics which have only been studied in certain limiting
cases \cite{and02,bob06}. 

In this paper we present a new, general
theoretical framework to compute the time-dependent transport
properties of SF hybrids. Compared to previous approaches
\cite{cue01,hue02,kop03}, our method allows one to find analytically
the time-dependent nonequilibrium (Keldysh) Green's functions near
spin-active interfaces. We apply the formalism to compute the spin
current-voltage characteristics of voltage-biased SFS junctions to
show that SFS structures can serve as highly tunable spin filters as
well as spin-current oscillators for spintronic applications.

We consider SFS point contacts to illustrate the dynamical
properties of spin-active Josephson junctions. The left ($1$) and
right ($2$) leads are assumed to be spin-singlet, $s$-wave
superconductors (S) in the clean limit. The ferromagnetic metallic
layer (F) is assumed to have a homogeneous magnetization fixed in
the $\hat{\vz}$ direction by anisotropy energy. We also assume the F
layer is specular and that its thickness is much less than the
superconducting coherence length and the mean free path in the
leads.
The contact is characterized by an S-matrix \cite{zha04}. Each
S-matrix component is a $2\times 2$ spin matrix, which for a fixed
quantization axis for spin, $\hat{\vz}$, is diagonal:
$S_{11}=S_{22}=\text{diag}\left(\sqrt{R_{\uparrow}}   e^{i\vartheta/2},
                                \sqrt{R_{\downarrow}} e^{-i\vartheta/2}\right)
                                \,,\,
 S_{21}=S_{12}=\text{diag}\left(i\sqrt{D_{\uparrow}}  e^{i\vartheta/2},
                                i\sqrt{D_{\downarrow}} e^{-i\vartheta/2}\right)$.
$D_{\alpha}$ is the transmission probability for normal state
conduction electrons with spin projection
$\alpha=\uparrow,\downarrow$, and $R_{\alpha}=1-D_{\alpha}$. In
general $D_{\uparrow}\neq D_{\downarrow}$, so the F layer acts like
a ``spin filter'';
$\vartheta$, referred to as the ``spin mixing angle'', is 
the relative phase between the transmitted (and reflected) spin up
and down electrons. Spin mixing at the F layer leads to rotation of
the direction of the spin polarization analogous to Faraday rotation
of linearly polarized photons \cite{tok88}.
Particle-hole symmetry relates the S-matrix for quasi-holes to that
for quasi-particles,
\underline{$S$}$_{ij}(\vp_{f_{\vert\vert}})=S_{ji}(-\vp_{f_{\vert\vert}})^{\mathrm{tr}}$
\cite{mil88}, which simplifies for junctions with inversion symmetry
to \underline{$S$}$_{ij}=S_{ij}$.
$D_{\alpha}$ and $\vartheta$ can be calculated from 
material parameters such as the exchange field, the Fermi wave vector mismatch (FWM), 
and the F layer thickness; see section III of Ref. \onlinecite{zha04} for 
detailed discussions on the interface S-matrix.
We emphasize that FWM plays an important role in FS 
junctions. For example, as shown in \cite{valls}, the subgap 
conductance of a FS junction is very sensitive to FWM and a 
nonmonotonic function of spin polarization. 
For transition metal ferromagnets such as Co 
and Ni, the transmission probabilities, $D_{\alpha}$, range from 0.2 to 0.9,
while $\vartheta$ can take any value between 0 and $\pi$, depending
on the F layer thickness \cite{sti96}.


For NFN junctions with normal metallic leads (N) under voltage bias
$V$, the spin current flowing through the junction is due to spin
filtering and linear in $V$, $I_\textsf{\tiny S}=G_\textsf{\tiny
S}\,V$. The spin conductance $G_\textsf{\tiny
S}\propto(D_{\uparrow}-D_{\downarrow})$ is independent of the
spin-mixing angle. Spin mixing, on the other hand, plays a central
role in transport through SFS junctions. For example, it can give
rise to $\pi$-junction behavior \cite{buz05,fog00}, as well as
complex features in the IV characteristics of the charge current,
i.e. the ``sub-harmonic gap structure'' (SGS) \cite{and02}.
For voltage-biased SFS contacts, the effective transmission
probability for spin up (down) quasiparticles in the superconducting
leads depends not only on $D_{\uparrow}$ and $D_{\downarrow}$, but
also on the spin-mixing angle ($\vartheta$) and the time-dependent
Josephson phase difference across the contact, $\phi(t)=2eVt/\hbar$.
The latter leads to inelastic scattering of quasiparticles of energy
$\epsilon$ by the spin-active interface, i.e. multiple Andreev
reflection (MAR) \cite{kla82,bra95}, which transfers quasiparticles
inelastically into side-band states with energies
$\epsilon_n\equiv\epsilon+n\hbar\omega$, with $\hbar\omega\equiv
eV$. As a result a time-dependent spin current flows across the
junction which has \emph{dc} and \emph{ac} components with
frequencies $2n\omega$, \be \mathbf{I}_\textsf{\tiny
S}(t)=\mathbf{I}_0+\sum_{n=1}^{\infty}(\mathbf{I}^c_n \cos 2n\omega
t+\mathbf{I}^s_n\sin 2n\omega t) \,. \ee
Our goal is to find the dependence of
$\{\mathbf{I}_0,\mathbf{I}^{c/s}_n\}$ on the bias voltage and
scattering parameters $\{D_{\alpha}, \vartheta\}$.

\par
Our method for computing time-dependent transport properties is
based on formulating the quasiclassical equations of nonequilibrium
superconductivity \cite{eli72} in terms of particle-hole coherence
functions and distributions functions, collectively known as the
{\sl Riccati amplitudes}. These functions obey Riccati-type
transport equations and uniquely determine the quasiclassical
Green's functions. The Riccati formulation is discussed in Refs.
\cite{esc00,zha04}; we follow the notation of these authors. For
superconductors in the clean limit the coherence function $\gamma^R$
($\tilde{\gamma}^R$) is the local probability amplitude for a hole
(electron) being converted into an electron (hole). The distribution
functions $x^K$ and $\tilde{x}^K$ describe the nonequilibrium
occupation of the particle and hole states.
Quasiparticle scattering near the contact, including spin filtering,
spin mixing and MAR, is described by a set of time-dependent
boundary conditions for the Riccati amplitudes. We have derived
these boundary conditions for spin-active interfaces in Ref.
\cite{zha04}. Here we present exact solutions for the boundary
conditions for the case of a constant bias voltage. The local
Green's functions at the contact are constructed from the Riccati
amplitudes and the spin current is calculated from the Keldysh
Green's function, $\hat{\g}^{K}$. We have checked that our method
reproduces known results for nonequilibrium charge transport through
spin-independent Josephson junctions \cite{kla82,bra95,ave95,cue96},
as well as the SFS structure studied in Ref. \cite{and02}.

We set the phase $\phi_1=0$ and $\phi_2=\phi(t)$ and work in a gauge
where $\phi(t)=2eVt/\hbar$. The corresponding Riccati amplitudes for
the two leads are (in the energy domain):
$\gamma^R_1(\epsilon',\epsilon)=2\pi\delta(\epsilon'-\epsilon)\gamma^R_1(\epsilon)$,
$x^K_1(\epsilon',\epsilon)=2\pi\delta(\epsilon'-\epsilon)x^K_1(\epsilon)$,
$\gamma^R_2(\epsilon',\epsilon)=2\pi\delta(\epsilon'-\epsilon+2\omega)\gamma^R_2(\epsilon-\omega)$,
and
$x^K_2(\epsilon',\epsilon)=2\pi\delta(\epsilon'-\epsilon)x^K_2(\epsilon+\omega)$.
The remaining Riccati amplitudes are obtained from those above by
symmetry relations: $\gamma_i^A=[\tilde{\gamma}_i^R]^{\dagger}$ and
$\tilde{q}(\hat{\vp}_f,\epsilon',\epsilon)\equiv
q^*(-\hat{\vp}_f,-\epsilon',-\epsilon)$ \cite{esc00}. Note that
$\gamma^R_2$ and $\tilde{\gamma}^R_2$ act as ladder operators in
energy space.

\begin{figure}
\includegraphics[width=3in]{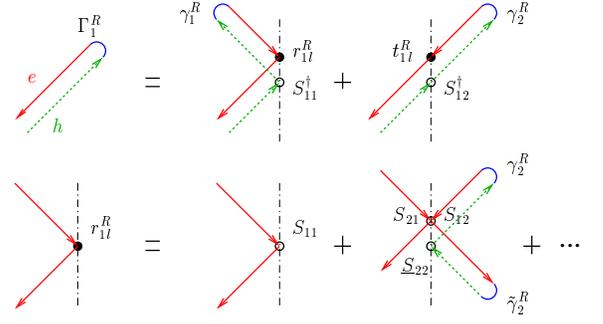}
\caption{\small Diagrammatic representation of
Eqs. (\ref{gamma_bc}). The solid (dash) arrows represent
electrons (holes), the vertical line indicates the
interface. Branch conversion amplitudes ($\gamma^R$ and
$\tilde{\gamma}^R$) are shown as semi-circles connecting
electron and hole lines, interface scattering amplitudes
($S_{ij}$) are shown as circles ($\circ$), and the effective
reflection and transmission amplitudes ($r^R_{1l}$ and $t^R_{1l}$)
are indicated by bullets ($\bullet$). Only the leading order multiple
scattering processes for $r^R_{1l}$ are shown.} \label{fig0}
\end{figure}

Consider a set of scattering trajectories $\{1i,1o,2i,2o\}$ near the
contact (``$i$''=incoming and ``$o$''=outgoing), c.f. Fig. 1 of Ref.
\cite{zha04}. Boundary conditions relate ``outgoing" and ``incoming"
Riccati amplitudes via the interface S-matrix. For example,
$\Gamma^R_1$ on trajectory $1i$ is given by,

\begin{eqnarray}
\Gamma^R_1 &=& r^R_{1l}\circ\gamma^R_1
\underline{S}^{\dagger}_{11} + t^R_{1l}\circ\gamma^R_2
\underline{S}^{\dagger}_{12},
\label{gamma_bc} \\
r^R_{1l} &=&
[S^{\dagger}_{11}-\beta_{21}\circ\beta_{22}^{-1}S^{\dagger}_{12}]^{-1},\;
t^{R}_{1l}=-r^R_{1l}\circ\beta_{21}\circ\beta_{22}^{-1}.\nonumber
\end{eqnarray}
where $\beta_{ij}\equiv
S^{\dagger}_{ij}-\gamma^R_j\underline{S}^{\dagger}_{ij}\circ\tilde{\gamma}^R_i$,
and the convolution is defined as $[x\circ y](\epsilon',\epsilon)=\int
d\epsilon''x(\epsilon',\epsilon'')y(\epsilon'',\epsilon)/ 2\pi$ in
energy domain. Equations (\ref{gamma_bc}) are the summation of
multiple scattering amplitudes represented by the diagrams
shown in Fig. \ref{fig0}.
%
%
Constructing $\hat{\g}^K$ from the Ricatti amplitudes
then leads us to evaluate the product,
\be
\mathrm{A}\equiv (1-\Gamma^R_1\circ\tilde{\gamma}^R_1)^{-1}\circ
r^R_{1l} =[\beta_{11}-\beta_{21}\circ\beta_{22}^{-1}\circ\beta_{12}]^{-1}
\,.
\ee
The ladder properties of $\gamma^R_2$ and
$\tilde{\gamma}^R_2$ lead to $\mathrm{A}(\epsilon',\epsilon)=\sum_m
2\pi\delta(\epsilon'-\epsilon-m\omega)\mathrm{A}_m(\epsilon)$, where
$\mathrm{A}_m(\epsilon)$ has the physical interpretation of the
transition amplitude for qusiparticles of energy $\epsilon$ in
lead 1 to be scattered into energy $\epsilon_m$. The calculation of
$\{\mathrm{A}_m\}$ requires inverting an infinite dimensional
tri-diagonal matrix in energy space, or equivalently solving the
three-term recurrence equation,
\begin{equation}
o_m\mathrm{A}_m+p_m\mathrm{A}_{m-2}+q_m\mathrm{A}_{m+2}=\delta_{m0}
\,,
\label{recur}
\end{equation}
where the coefficients are defined by
\begin{eqnarray}
o_m&=&\beta_{11}(\epsilon_m)
-S^{\dagger}_{21}\beta_{22}^{-1}(\epsilon_{m+1})S^{\dagger}_{12}
\nonumber\\
&\;&-p_m(\epsilon)[{S}^{\dagger}_{12}]^{-1}\gamma^R_2(\epsilon_{m-1})\underline{S}^{\dagger}_{12}
\tilde{\gamma}^R_1(\epsilon_{m})\,,
\\
p_m&=&\gamma^R_1(\epsilon_m)\underline{S}^{\dagger}_{21}\tilde{\gamma}^R_2(\epsilon_{m-1})
\beta_{22}^{-1}(\epsilon_{m-1}){S}^{\dagger}_{12}\,,
\\
q_m&=&S^{\dagger}_{21}\beta_{22}^{-1}(\epsilon_{m+1})\gamma^R_2(\epsilon_{m+1})
\underline{S}^{\dagger}_{12}\tilde{\gamma}^R_1(\epsilon_{m+2})\,.
\end{eqnarray}
Equation (\ref{recur}) is solved by introducing transfer matrices, $Z^{\pm}_m$, such
that $\mathrm{A}_{m}=Z^+_m\mathrm{A}_{m-2}$ for $m>0$ and
$\mathrm{A}_{m}=Z^-_m\mathrm{A}_{m+2}$ for $m<0$. The $Z^{\pm}_m$ can be expressed as
matrix continued fractions $Z_m^+=[o_m+q_mZ_{m+2}^+]^{-1}(-p_m)$,
$Z_m^-=[o_m+p_mZ_{m-2}^-]^{-1}(-q_m)$, and evaluated using standard procedures
\cite{ris80}. The resulting $Z^{\pm}_m$ allow us to compute $\mathrm{A}_m$ for even $m$
starting from $\mathrm{A}_0=[o_0+p_0Z_{-2}^{-}+q_0Z^+_{2}]^{-1}$. The
amplitudes for odd $m$  vanish identically.

The angle- and energy-resolved spin-current evaluated to the left of
the contact is given by $ \mathbf{J}^l_s(\hat{\vp}_f)=
g(\hat{\vp}_f)\Tr{\vsigma\hat{\tau}_3(\hat{\g}^K_{1i}-\hat{\g}^K_{1o})}/2\pi
i$, where the trace is over particle-hole and spin space,
$g(\hat{\vp}_f)=N_f\Area{\hbar\over 2}(\vv_f\cdot\hat{\vn})$, $N_f$
is the density of states at the Fermi level, $\vv_f$ is the Fermi
velocity, $\mathcal{A}$ is the contact area, and $\hat{\vn}$ is the
normal to the junction interface. We evaluate the current by
expressing $\hat{\tau}_3(\hat{\g}^K_{1i}-\hat{\g}^K_{1o})$ in terms
of the local Riccati amplitudes and carry out the trace over
particle-hole space. We obtain a sum of 8 terms; e.g. the
contributions from electron-like quasiparticles from lead 1, which
are proportional to $x^K_1$, are $\mathrm{A}\circ x^K_1\circ
\mathrm{A}^{\dagger}$ and $\tilde{\gamma}^R_1\circ \mathrm{A}\circ
x^K_1\circ\mathrm{A}^{\dagger}\circ \gamma^A_1$. Similarly,
contributions from $\tilde{x}^K_1$, $x^K_2$, and $\tilde{x}^K_2$ are
expressed as convolutions with transition amplitudes B, C, D, and E,
which can all be expressed in terms of A [see Eq. (\ref{kernelB})].
We evaluate the convolution in the energy domain, then Fourier
transform to obtain the trajectory-resolved dynamical spin current,
\begin{eqnarray}
\mathbf{J}^l_s(\hat{\vp}_f;t) = g(\vp_f)\,
\mathrm{Re}\sum_{k,m}e^{i2k\omega t} \int{d\epsilon\over
2\pi}\mathrm{Sp} \left[\vsigma\mathrm{K}_{mk}(\epsilon)\right] ,
\label{spin-current_spectrum}
\end{eqnarray}
where $\mathrm{Sp}$ is a trace over spin space. The kernel
$\mathrm{K}_{mk}(\epsilon)$ consists of 8 terms which describe
interference between inelastic scattering
$\epsilon\rightarrow\epsilon_{2m}$ and
$\epsilon\rightarrow\epsilon_{2(m+k)}$:
{\small
\begin{eqnarray}
\mathrm{K}_{mk}(\epsilon)&=&\mathrm{H}_{mk}(\epsilon)
+\tilde{\gamma}^R_1(\epsilon_{2m})\mathrm{H}_{mk}(\epsilon)[\tilde{\gamma}^R_1(\epsilon_{2(m+k)})]^{\dagger}
\nonumber\\
&-&\mathrm{D}_{2m}\tilde{x}^K_1(\epsilon)
\mathrm{D}^{\dagger}_{2(m+k)}
-\mathrm{E}_{2m}\tilde{x}^K_1(\epsilon)
\mathrm{E}^{\dagger}_{2(m+k)},
\nonumber\\
\mathrm{H}_{mk}(\epsilon)&=&
\mathrm{A}_{2m}x^K_1(\epsilon)\mathrm{A}^{\dagger}_{2(m+k)}
+\mathrm{B}_{2m}x^K_2(\epsilon_{1})\mathrm{B}^{\dagger}_{2(m+k)}
\nonumber\\
&-&\mathrm{C}_{2m}\tilde{x}^K_2(\epsilon_{-1})\mathrm{C}^{\dagger}_{2(m+k)},
\nonumber\\
\mathrm{B}_{m}(\epsilon)&=&
[\mathrm{A}_{m-2}(\epsilon_{2})\gamma^R_1(\epsilon_2)\underline{S}^{\dagger}_{21}\tilde{\gamma}^R_2(\epsilon_{1})
-\mathrm{A}_{m}(\epsilon)S^{\dagger}_{21}]\beta^{-1}_{22}(\epsilon_{1}),
\nonumber \\
\mathrm{C}_{m}(\epsilon)&=&\{\mathrm{D}_m(\epsilon)\underline{S}_{12}
-[\mathrm{A}_{m+2}(\epsilon_{-2})S^{\dagger}_{11}
\nonumber\\
&+&\mathrm{B}_{m+2}(\epsilon_{-2})S^{\dagger}_{12}
]{S}_{12}\gamma^R_2(\epsilon_{-1})\}
\tilde{\beta}^{-1}_{22}(\epsilon_{-1}),
\nonumber\\
\mathrm{D}_{m}(\epsilon)&=&
\mathrm{A}_m(\epsilon)\gamma^R_1(\epsilon)\underline{S}^{\dagger}_{11}
+\mathrm{B}_{m+2}(\epsilon_{-2})\gamma^R_2(\epsilon_{-1})\underline{S}^{\dagger}_{12},
\nonumber\\
\mathrm{E}_m(\epsilon)&=&\tilde{\gamma}_1(\epsilon_m)[\mathrm{A}_m(\epsilon)S^{\dagger}_{11}
+\mathrm{B}_m(\epsilon)S^{\dagger}_{12}]\tilde{\gamma}^{-1}_1(\epsilon).
\label{kernelB}
\end{eqnarray}
}
Thus, the calculation of the spin current is reduced to the
calculation of the \{A$_m$\}. Integration over all contributing
trajectories yields the total spin current on the left side of the
junction,
$\mathbf{I}^{l}_s(t)=\langle\mathbf{J}^{l}_s(\hat{\vp}_f,t)\rangle$,
where $\langle ...\rangle=\int^{\pi/2}_0 d\psi\sin (2\psi) (...)/4$,
and $\psi=\arccos(\hat{\vv}_f\cdot \hat{\vn})$.
%
%
Eqs. (\ref{spin-current_spectrum})-(\ref{kernelB}) are our central
results. They are valid for very general contact scattering matrices.

In general spin-active junctions do not conserve the spin current.
However, for the SFS point contacts described above, only the $\vz$
component of spin current is nonzero and it is conserved across the
contact, $\mathbf{I}_s(t)=I_s(t)\hat{\vz}$.
Furthermore, the suppression of order parameter near the point
contact, as well as spin accumulation in the leads, are negligible.
Thus, $\gamma^R_j(\epsilon)$, $\tilde{\gamma}^R_j(\epsilon)$,
$x^K_j(\epsilon)$, and $\tilde{x}^K_j(\epsilon)$ can be evaluated in
terms of their bulk values in the corresponding leads \cite{zha04}.
For the results reported below we neglected the directional
dependence of $D_{\alpha}$ and $\vartheta$ and assumed an inelastic
scattering rate of $10^{-4}\Delta$. The IV characteristics, when
normalized by the temperature-dependent gap $\Delta(T)$, depend only 
weakly on the temperature, so we only present results for $T=0$.
%
\begin{figure}
\includegraphics[width=3in]{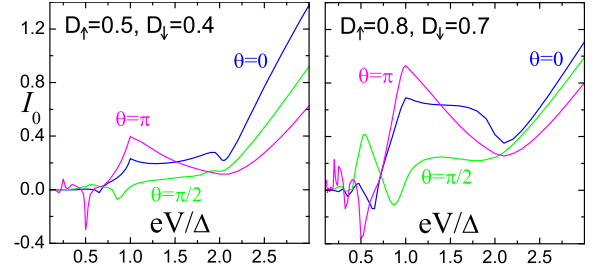}
\caption{\small The \emph{dc} spin current (in units of $G_\textsf{\tiny S}\Delta/e$)
         as a function of voltage bias $V$ at zero temperature, for
         different transmission probabilities,
         $D_{\uparrow}$ and $D_{\downarrow}$, and
         spin-mixing angles, $\vartheta=0$
         (blue), $\pi/2$ (green), $\pi$ (red).}
\label{fig1}
\end{figure}

Figure \ref{fig1} shows the \emph{dc} component of spin current at
zero temperature for two sets of spin dependent transparencies, and
three values of the spin mixing angle. Consider first the case of
pure spin filtering, i.e. $\vartheta=0$. For $eV> 2\Delta$, the spin
current is always less than the normal state value,
$G_{\textsf{\tiny S}}V$, where the spin conductance $G_\textsf{\tiny
S}= N_f v_f{\hbar\over 2}\mathcal{A}\langle
D_{\uparrow}-D_{\downarrow}\rangle$. The SGS of the spin current
shows anomalies at voltages, $V_n=2\Delta /ne$, which are more
pronounced in higher transmission junctions. The structures
associated with even and odd harmonics are different: the spin
current is a local minimum near $V_{2m+1}$, but a local maximum near
$V_{2m}$.
The SGS for the spin current is in sharp contrast with that of the
charge current \cite{and02}. The deficit in
spin current at high voltages, as well as the even-odd effect in the
SGS reflect the quasiparticle origin of the spin current. Multiple Andreev
reflection processes starting at subgap energies contribute significantly
to the charge current, but much less so to the spin current.
Consider a scattering event where an incoming spin-up electron-like
quasiparticle of energy
$\epsilon\lesssim-\Delta$ in lead 1 climbs up the MAR ladder to
energy $\epsilon_n\gtrsim\Delta$ after $n-1$ Andreev
reflections ($n\sim 2\Delta/eV$) and leaks into the leads. This
process transfers charge $ne$ to lead 2, but only
spin angular momentum $\hbar/2$, and only if $n$ is odd. The $n^{\text{\small th}}$
order MAR process has an onset voltage, $V_n$. Opening up new
transmission channels at $V_{2m+1}$ leads to increase of the spin
current, while the onset of back scattering at $V_{2m}$ leads to
decrease in spin current. For larger contact transparencies
contributions from higher-order MAR processes become relevant,
and more detailed structures in the IV curve develop at low
voltages, as are evident in the right panel of Fig. \ref{fig1}.

The inclusion of spin mixing leads to interface bound states at
energies $\Delta\cos(\vartheta/2)$ [spin up] and
$-\Delta\cos(\vartheta/2)$ [spin down] \cite{fog00}.
Qualitatively, this structure of the local density of states
(LDOS) affects the spin transport in two ways. First, the LDOS
for continuum states near the gap edge is depleted. This
leads to suppression of the spin current at high
voltages; also the SGS at the onset voltages for MAR processes
is less pronounced. Second, scattering
events with MAR trajectories which coincide with the bound states
exhibit resonant transmission and dominate the spin current.
Thus, in relatively low-transmission junctions the SGS is displaced from $V_n$
to voltages near $\Delta(1+\cos(\vartheta/2))/n$ \cite{and02}. Since the resonances
for spin up and spin down quasiparticles occur at different voltages,
the \emph{dc} spin current oscillates with voltage, even reversing direction
in junctions with moderate to high transparency. For
$\vartheta=\pi$, the bound states are degenerate and located at the Fermi level,
and consequently the SGS at $V_n$ is recovered.
%
\begin{figure}
\includegraphics[width=3in]{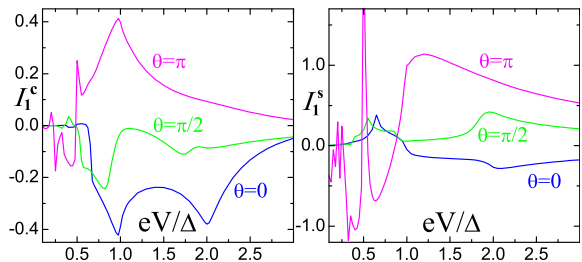}
\caption{\small Left (right): The cosine (sine) components of the
\emph{ac} spin current (in units of $G_\textsf{\tiny S}\Delta/e$) at
$T=0$ for $\vartheta=0,\pi/2,\,\pi$,
$D_{\uparrow}=0.5$, and $D_{\downarrow}=0.4$.}
\label{fig2}
\end{figure}

Figure \ref{fig2} shows the cosine and sine part of the first
Fourier component of the \emph{ac} spin current at zero temperature
for intermediate transparencies: $D_{\uparrow}=0.5,D_{\downarrow}=0.4$.
Note that the magnitudes of $I_1^c$ and $I_1^s$ are comparable to
that of the \emph{dc} current. However, higher order Fourier
components of the \emph{ac} spin current are generally smaller in
magnitude, but readily calculated from Eqs.
(\ref{spin-current_spectrum})-(\ref{kernelB}).
As in the case of the
\emph{dc} current, the structures in $I_1^c(V)$ and $I_1^s(V)$ are
related to onsets and resonances associated with MAR processes.
Resonant transmission leads to dramatic variations and sharp peaks
in the \emph{ac} spin current at voltages less than $2\Delta$; e.g.
for $\vartheta=\pi$ the sine-component dominates the response near
$eV/\Delta=0.5$ and reverses sign over a narrow voltage range,
$e\delta V/\Delta\approx 0.1$.
\par
In conclusion, we calculated the dynamical spin current in
voltage-biased spin-active Josephson contacts. The dependence of the
spin current on the voltage bias exhibits nonlinearities and
resonant structures that are in striking contrast to the IV
characteristics of the charge current, but which can be understood
in terms of spin-filtering, spin-mixing and MAR.
The nonlinear spin IV characteristics may be used to control the
magnitude and the direction of the spin current via bias voltage,
and may be of interest for superconducting spintronic devices.
\par
We thank Tomas L\"ofwander for discussions on \textsl{ac} Josephson
effects.

%

\end{document}